\input amstex
\input xy
\xyoption{all}
\input epsf
\documentstyle{amsppt}
\document
\magnification=1200  
\NoBlackBoxes
\nologo
\hoffset=1.5cm
\voffset=1truein
\def\r{\roman}

\def\d{\partial}

\def\Q{\bold{Q}}
\def\Z{\bold{Z}}
\def\R{\bold{R}}
\def\C{\bold{C}}



\vskip1cm

\pageheight{16cm}


\bigskip
 \centerline{\bf SYMBOLIC DYNAMICS, MODULAR CURVES,}
 
  \medskip
 
 \centerline{\bf AND  BIANCHI IX   COSMOLOGIES}
 
  \bigskip

\centerline{ Yuri I. Manin${}^1$, Matilde Marcolli${}^2$}

\medskip

\centerline{\it ${}^1$Max--Planck--Institut f\"ur Mathematik, Bonn, Germany,}
\smallskip
\centerline{\it ${}^2$California Institute of Technology, Pasadena, USA}
\vskip1cm

\hfill{To Vadim Schechtman, most cordially}

\bigskip
{\it ABSTRACT. It is well known that the so called Bianchi IX spacetimes
with $SO(3)$--symmetry   in a neighbourhood of the Big Bang exhibit a chaotic  behaviour
of  typical trajectories in the backward movement of time. This behaviour
(Mixmaster Model of the Universe) can be  encoded
by the shift of two--sided continued fractions.

Exactly the same shift encodes the sequences of intersections 
of hyperbolic geodesics with purely imaginary axis 
in the upper complex half--plane, that is geodesic flow on an appropriate modular surface. 

A physical interpretation
of this coincidence was suggested in  [MaMar14]: namely, that Mixmaster chaos
is an approximate description of the passage from a hot quantum Universe at the Big Bang
moment to the cooling classical Universe.   Here we discuss and 
elaborate this suggestion, looking at the Mixmaster Model from the
perspective of the second class of Bianchi IX spacetimes: those  with $SU(2)$--symmetry
(self--dual Einstein metrics). We also extend it to the more general context related
to Painlev\'e VI equations.}

\bigskip

\centerline{\bf 1. Introduction, background and notation}

\medskip

{\bf 1.1. Plan of the paper.}  The  Mixmaster Model
of the early Universe with $SO(3)$--symmetry in a neighbourhood of the Big Bang 
predicts a chaotic  behaviour
of  ``typical" trajectories (in the backward movement of time)  encoded
by the shift of two--sided continued fractions: cf. [KLKhShSi85], [BoNo73], 
[May87], and references therein.

\smallskip

The same shift encodes the sequence of intersections  with purely imaginary axis 
of hyperbolic geodesics in the upper complex half--plane, see [Se85]. 

\smallskip

This coincidence invites a closer attention, because the accepted mathematical
source of  the classical Mixmaster chaos involves the behaviour of separatrices
on the real boundary of the respective dynamical system (cf. [Bo85]).
Geometry of these separatrices and approximate dynamics that it encodes
are not visibly  related  to hyperbolic geodesics.

\smallskip

A physical interpretation
of this coincidence was suggested in  [MaMar14].  Here we discuss and 
elaborate this suggestion, looking at the Mixmaster model from the
perspective of Bianchi IX model with $SU(2)$--symmetry.

\smallskip

More precisely, according to [MaMar14],
the Mixmaster ``classical chaos'' should be considered as {\it an approximation
to an unknown quantum description} of the transition from the infinitely hot quantum Universe
at the moment of Big Bang to the cooling Universe gradually fitting  a classical 
Einsteinian model.  Time axis at the moment of Big Bang is purely imaginary,
and it becomes real during the observable history of Universe.

\smallskip

We argued  that a mathematical model of
such a transition explaining Mixmaster chaos consists in inverse Wick rotation of time axes {\it mediated by a move of
time along  random geodesics in the complex hyperbolic half--plane} or rather
its appropriate modular quotient. This passage to the   modular quotient
was critically important  for our argument. It was suggested by two initially disjoint
evidences. The first one was P.~Tod's  remark that a conformal version of
cosmological time in the Friedman--Robertson--Walker models 
has a natural structure of the elliptic integral (cf. [MaMar14], sec.~4.2).
The second evidence was a well known formal coincidence
of two encodings: of Kasner's trajectories, on the one hand, and of
hyperbolic geodesics with ideal ends, on the other hand.

\smallskip
In this paper we develop and present further details of this picture.
Namely, we now look at such a transition from the side
of ``gravitational instantons'' that is, self--dual Einstein spacetimes 
 with $SU(2)$--symmetry.
Many such spacetimes have a natural complexification,
in particular, time axis can be extended to the complex half--plane, 
whereas the instantons themselves are defined by restricting time
to the imaginary semi--axis.

\smallskip

Following the behaviour of the respective models along oriented  geodesics in time connecting
imaginary half--axis with  real half--axis, we get the new
aspect of the Mixmaster picture. This is the main content of this note.

\smallskip

{\it Structure of the paper.} In the remaining part of  section 1, we introduce some basic notation and constructions.

\smallskip
Sec.~2 compares (and shows a satisfying agreement) the sequences of Kasner eras
in the classical Mixmaster models with sequences of geodesic distances between
consecutive intersection of a geodesic with sides of the Farey tessellation.
Finally, in sec.~3 and 4 we study an ``instanton analogue'' of the sequences of Kasner
solutions determining chaotic behaviour in the classical Mixmaster model. 

\medskip

{\bf 1.2. Continued fractions.} We denote by $\Z$, resp. $\Z_+$, the set of integers,
resp. positive integers; $\Q$, resp. $\R$  is the field of rational, resp. real numbers.
For $x\in \R$, we put $[x]:=\r{max}\,\{m\in\Z\,|\,m\le x\}.$
\smallskip
Irrational numbers $x>1$ admit the canonical infinite continued fraction representation
$$
x= k_0+\frac{1}{k_1+\frac{1}{k_2+\dots}}=: [k_0,k_1,k_2,\dots ],\ k_s\in \Z_+
\eqno(1.1)
$$
in which $k_0:=[x]$, $k_1=[1/{x}]$ etc.
Notice that our convention differs from that of  [KLKhShSi85]: their
$[k_1,k_2, \dots]$ means our  $[0, k_1,k_2, \dots]$.

\medskip

{\bf 1.3. Transformation $T$.} The (partial) map
$\widetilde T:\, [0,1]^2\to [0,1]^2$ is defined by
$$ 
\widetilde T: (x,y) \mapsto \left( \frac{1}{x} - \left[ \frac{1}{x}
\right], \frac{1}{y+[1/x]} \right), 
\eqno(1.2)
$$
If both coordinates $(x,y)\in [0,1]^2$ are irrational (the complement is a subset of measure
zero), we have for uniquely defined $k_s\in \Z_+$:
$$
x=[0,k_0,k_1,k_2, \dots ], \  y=[0,k_{-1}, k_{-2}, \dots ].
$$
Then
$$
\frac{1}{x} -\left[ \frac{1}{x}\right] = [0,k_1,k_2,\dots ], \quad \frac{1}{y+[1/x]} = \frac{1}{k_0+y}=[0,k_0,k_{-1},k_{-2}, \dots ].
$$
On this subset, $\widetilde T$ is bijective and has invariant density
$$
\frac{dx\,dy}{\r{ln}\,2\cdot (1+xy)^2}
$$
(cf. [May87]).
\smallskip
Thus we may and will bijectively encode irrational pairs $(x,y)\in [0,1]^2$ by doubly
infinite sequences
$$
(k):=[\dots k_{-2},k_{-1},k_0,k_1, k_2,\dots ], k_i\in \Z_+
$$
in such a way that the map $\widetilde T$ above becomes the shift of such a sequence  denoted $T$:
$$
T(k)_s=k_{s+1}.
\eqno(1.3)
$$

\medskip

{\bf 1.4. Continued fractions and chaos in Einsteinian Bianchi IX type models.}
Bianchi classified metric space-times with a Lie group action transitive on space sections. In particular 4--dim
Bianchi IX models of space--time can be of two types: with the symmetry group $SO(3)$
or else $SU(2)$. In the first case, metric has Minkowski's signature, whereas
in the second case it is Riemannian. In sec.~1 we survey the now classical results
about chaotic behaviour in the $SO(3)$--case (Mixmaster Universe) and prepare ground
for the treatment of $SU(2)$--models.
Sec. 2 and 3 are dedicated to the $SU(2)$--case.

\smallskip
Consider the real circle defined in $\R^3$ by equations
$$
p_a+p_b+p_c=1, \quad p_a^2+p_b^2+p_c^2=1.
\eqno(1.4)
$$
Each point of this circle defines a 4--dimensional space--time with metric of
Minkowski signature $dt^2-a(t)dx^2-b(t)dy^2-c(t)dz^2$ with {\it scaling factors} $a,b,c$:
$$
a(t) =t^{p_a},\  b(t) =t^{p_b},\  c(t) =t^{p_c}, t >0.
$$
Such a metric is called the Kasner metric with exponents $(p_a, p_b, p_c)$.

\smallskip

Any point $(p_a,p_b,p_c)$ can be obtained by choosing a unique $u\in [1,\infty ]$,
putting
$$
p_1^{(u)}:= -\frac{u}{1+u+u^2}\in [-1/3, 0],\
p_2^{(u)}:= \frac{1+u}{1+u+u^2}\in [0, 2/3],\
$$
$$
p_3^{(u)}:= \frac{u(1+u)}{1+u+u^2}\in [2/3, 1]
\eqno(1.5)
$$
and then rearranging the  exponents $p_1^{(u)}\le p_2^{(u)}\le p_3^{(u)}$ by a bijection $(1,2,3)\to (a,b,c)$.

\smallskip

The main result of a series of physical papers dedicated  to the Mixmaster Universe
can be roughly summarized  as follows.
\smallskip
A ``typical'' solution $\gamma$  of Einstein equations (vacuum, but also with various energy momentum tensors) with $SO(3)$--symmetry of the Bianchi IX type,
followed from  an arbitrary (small) value $t_0>0$ in the reverse time direction $t\to +0$, oscillates close to a sequence of Kasner type solutions. (See subsection 2.3 below qualifying the use
of  adjective ``typical'' in this context).

\smallskip

Somewhat more precisely, introduce the local logarithmic time $\Omega$  along this trajectory with inverted orientation. Its differential is
$d\Omega:=-\dfrac{dt}{abc}$, and the time itself is  counted from an arbitrary but fixed moment.
Then $\Omega \to +\infty$ approximately as $-\roman{log}\,t$ as $t\to +0$, and we have the following picture. 
\smallskip

 As $\Omega \cong - \roman{log}\,t \to +\infty$,
 a ``typical'' solution $\gamma$ of the Einstein equations
determines a sequence of infinitely increasing moments 
$\Omega_0<\Omega_1< \dots <\Omega_n< \dots$
and a sequence of irrational real numbers $u_n\in (1, +\infty ),\  n=0,1,2,\dots $.

\smallskip

The time semi--interval $[\Omega_n,\Omega_{n+1})$ is  called the {\it $n$--th Kasner era} (for the trajectory $\gamma$).     
Within the $n$--th era, the evolution of $a,b,c$ is approximately
described by several consecutive 
Kasner's formulas.  Time intervals where  scaling powers $(p_i)$  are (approximately)  constant
are called {\it Kasner's cycles.}

\smallskip
 The evolution in the $n$--th era starts at time $\Omega_n$ with a certain value $u=u_n>1$
which determines the sequence of  respective scaling powers during the first cycle (1.5):
$$
p_1=-\frac{u}{1+u+u^2},\ p_2=\frac{1+u}{1+u+u^2},\ p_3=\frac{u(1+u)}{1+u+u^2}
$$
The next cycles inside the same era start with values $u=u_n-1$, $u_n-2, \dots$,
and scaling powers (1.5) corresponding to these numbers, rearranged corresponding
to a bijection $(1,2,3)\to (a,b,c)$ which is in turn identical to the previous one, or interchanges $b$ and $c$
(see [MaMar02]).

\smallskip

 After $k_n:=[u_n]$ cycles inside the current era, a jump to the next era comes,
with parameter
$$
u_{n+1}=\frac{1}{u_n-[u_n]}.
\eqno(1.6)
$$

\smallskip

Moreover, ensuing encoding of $\gamma$'s and respective sequences $(u_i)$'s
by continued fractions (1.1) of real irrational numbers $x>1$ is bijective on the set of full measure.

\smallskip

Finally, when we want to include into this picture also the sequence of logarithmic times
$\Omega_n$ starting new eras, we naturally pass to the two--sided continued fractions 
and the transformationn $T$. See some details in the next section.

\medskip

{\bf  1.5. Doubly infinite sequences and modular geodesics.} Let $H:= \{z\in \C,\ \r{Im}\, z >0\}$
be the upper complex half--plane with its Poincar\'e metric $|dz|^2/ |\r{Im}\,z |^2$.
Denote also by $\overline{H}:= H\cup \{\Q\cup\{\infty\}\}$ this half--plane completed with cusps.

\smallskip

The vertical lines $\r{Re}\,z=n, n\in \Z$, and semicircles in $\overline{H}$ connecting pairs of finite cusps
$(p/q,p^{\prime}/q^{\prime})$ with $pq^{\prime}-p^{\prime}q =\pm 1$, cut $\overline{H}$ into the union of geodesic ideal triangles
which is called the {\it Farey tessellation.}

\smallskip

Following [Se85], consider the set of {\it oriented} geodesics $\beta$'s in $H$ with ideal irrational endpoints in 
$\R $.
Let $\beta_{-\infty}$, resp. $\beta_{\infty}$ be the initial, resp. the final point of $\beta$.
Let $B$ be the set of such geodesics with $\beta_{-\infty}\in (-1,0)$, $\beta_{\infty}\in (1,\infty)$. Put
$$
\beta_{-\infty} = -[0,k_0,k_{-1},k_{-2},\dots ],\quad  \beta_{\infty} = [k_1,k_2,k_3,\dots ], \quad k_i\in \Z_+,
\eqno(1.7)
$$
and encode $\beta$ by the doubly infinite continued fraction
$$
[\dots k_{-2}, k_{-1}, k_0, k_1, k_2,\dots ] .
\eqno(1.8)
$$
The geometric meaning of this encoding can be explained as follows. Consider the intersection point  $x=x(\beta )$ 
of $\beta$ with the imaginary semiaxis in $H$. Moving along $\beta$ from $x$ to $\beta_{\infty}$,
one will intersect an infinite sequence of Farey triangles. Each triangle is entered through a side
and left through another side, leaving the ideal intersection point (a cusp) of these sides either
to the left, or to the right. Then the infinite word in the alphabet $\{L,R\}$ encoding the consecutive positions
of these cusps wrt $\beta$ will be $L^{k_1}R^{k_2}L^{k_3}R^{k_4}\dots $ Similarly, moving from  $\beta_{-\infty}$
to $x$, we will get the word  (infinite to the left)  $ \dots L^{k_{-1}}R^{k_0}$.

\smallskip

We can enrich  the new notation  $\dots L^{k_{-1}}R^{k_0}L^{k_1}R^{k_2}L^{k_3}R^{k_4}\dots $
(called {\it cutting sequence} of our geodesic in [Se85])
by inserting between the consecutive powers of $L,R$ notations for 
the respective intersection points of $\beta$ with the sides of Farey triangles. So
$x_0:=x=x(\beta)$ will be put between $R^{k_0}$ and $L^{k_1}$, and generally we
can imagine the word
$$
\dots L^{k_{-1}}x_{-1}R^{k_0}x_0L^{k_1}x_1R^{k_2}x_2L^{k_3}x_3R^{k_4}\dots 
\eqno(1.9)
$$
We will essentially use this enrichment in the next section.
\bigskip

\centerline{\bf 2. Hyperbolic billiard, geodesic distance, and cosmological time}

\medskip

{\bf 2.1. Hyperbolic billiard.} We will first present a version of the geometric description of geodesic
flow:  an equivalent
dynamical system which is the triangular hyperbolic billiard with infinitely distant corners (``pockets'').

\smallskip
Here we use the term ``hyperbolic'' in order to indicate that sides (boards) of the billiard and trajectories
of the ball (``particle'') are geodesics with respect to the hyperbolic metric of constant curvature $-1$ of the billiard table. This is not the standard meaning of the hyperbolicity in this context,
where it usually refers to non--vanishing Lyapunov exponents.

\medskip

{\bf 2.2. Proposition.} {\it a) All hyperbolic triangles of the Farey tessellation of $\overline{H}$ are isomorphic
as metric spaces.

\smallskip

b) For any two closed triangles having a common side there exists unique metric isomorphism of them identical
along this side. It inverts orientation induced by $H$. Starting with the basic triangle $\Delta$ with vertices 
$\{0,1, i\infty \}$ and  consecutively using these identifications, one can unambiguously define the map
$b:\, \overline{H}\to \Delta$.

\smallskip

c) Any oriented geodesic on $H$ with irrational end--points in $\bold{R}$ is sent by the map $b$
to a billiard ball trajectory on the table $\Delta$ never hitting corners.}

\medskip

All this is essentially well known. 

\smallskip

It is also worth noticing that although all three sides  of $\Delta$ are of infinite length,
this triangle is {\it equilateral} in the following sense: there exists a  group $S_6$ of
hyperbolic isometries of $\Delta$ acting on vertices by arbitrary permutations.
This group has a unique fixed point $\rho := \roman{exp} (\pi i/3)$ in $\Delta$, {\it the centroid} of $\Delta$.

\smallskip

In fact, this group is generated by two isometries: $z\mapsto 1-z^{-1}$ and symmetry
with respect to the imaginary axis.

\smallskip

Three finite geodesics connecting the centre $\rho$ with points $i, 1+i, \frac{1+i}{2}$ respectively,
subdivide $\Delta$ into three geodesic quadrangles, each having one infinite (cusp)  corner.
We will call these points  {\it centroids} of the
respective sides of $\Delta$, and  the geodesics $(\rho ,i)$ etc. {\it medians} of $\Delta$.

\smallskip
Each quadrangle is the fundamental domain for $PSL (2,\bold{Z})$.

\medskip

{\bf 2.3. Billiard encoding of oriented geodesics.} Consider the first stretch of the geodesic $\beta$
encoded by  (1.9) that
starts at the point $x_0$ in $(0,i\infty )$. If $k_0=1$,the ball along $\beta$ reaches the opposite side
$(1,i\infty )$ and gets reflected to the third side $(0,1)$. If $k_0=2$, it reaches the opposite side, then returns to the initial side $(0, i\infty )$, and only afterwards gets reflected to $(0,1)$.

\smallskip

More generally, the ball always spends $k_0$ unobstructed stretches of its trajectory between 
$(0,i\infty )$ and $(1,i\infty )$, but then is reflected to $(0,1)$ either from $(1, i\infty)$
(if $k_0$ is odd), or from $(0,i\infty )$ (if $k_0$ is even). We can encode this sequence of stretches
by the formal word $\infty^{k_0}$ showing exactly how many times the ball is reflected
``in the vicinity'' of the pocket $i\infty$, that is, does not cross any of the medians.

\smallskip

A contemplation will convince the reader that this allows one to define an alternative encoding
of $\beta$ by the double infinite word in {\it three letters }, say $a,b,c$, serving as names  
of the vertices $\{0,1,i\infty\}$. 

\smallskip

{\bf 2.4. Kasner's eras in logarithmic time and doubly infinite continued fractions.}
Now we will explain,  how the double infinite continued fractions
enter the Mixmaster formalism when we want to mark the consecutive Kasner eras
upon the $t$--axis, or rather upon the $\Omega$--axis, where $\Omega := -\r{log}\,\int dt/abc$

\smallskip

In the process of construction, these continued fractions will also come with their
enrichments, and the first new result of this note will compare
this enrichment with the one described by (1.9).

\smallskip

We start with fixing a ``typical'' space--time $\gamma$ whose evolution with $t\to +0$ undergoes
(approximately) a series of Kasner's eras described by a continued
fraction $[k_0,k_1,k_2, \dots ]$, where $k_s$ is the number of Kasner's cycles within
$s$--th era $[\Omega_s,\Omega_{s+1})$. We have enriched this encoding by introducing parameters
$u_s$ which determine the Kasner exponents within the first cycle of the era number $s$
by (1.5). A further enrichment comes with putting these eras on the $\Omega$--axis.
According to [KLKhShSi85], [BoNo73], [Bo85], if one defines the sequence of numbers $\delta_s$
from the relations 
$$
\Omega_{s+1}=[1+\delta_sk_s(u_s+1/ \{u_s\})]\Omega_s,
$$
then complete information about these numbers can be encoded by the extension to the left
of our initial continued fraction:
$$
[\dots , k_{-1},k_0,k_1,k_2,\dots ]
\eqno(2.1)
$$
in such a way that
$$
\delta_s=x_s^+/(x_s^+ +x_s^-)
$$
where
$$
x_s^+=[0,k_s,k_{s+1}, \dots ], \quad  x_s^- =[0,k_{s-1},k_{s-2}, \dots ] .
\eqno(2.2)
$$
\medskip

{\bf 2.5. Theorem.} {\it Let a  ``typical'' Bianchi IX Mixmaster Universe be encoded by the double--sided
sequence (2.1).
Consider also the respective geodesic in $H$ with its enriched encoding (1.9). 

\smallskip

Then we have ``asymptotically''  as $s\to \infty$, $s\in \Z_+$:
$$
\r{log}\, \Omega_{2s}/ \Omega_0 \simeq  2 \sum _{r=0}^{s-1} \r{dist}\, (x_{2r},x_{2r+1}),
\eqno(2.3)
$$
where $\r{dist}$ denotes the hyperbolic distance between the consecutive intersection points 
of the geodesic with sides of the Farey tesselation
as in (1.9).}

\medskip

{\bf Proof.} According to the formulas (5.1) and (5.5) in  [KLKhShSi85], and our notation (2.2), we have
$$
\r{log}\, \Omega_{2s}/ \Omega_0 \simeq
-\sum_{p=1}^{2s}\r{log}\ (x^+_px^-_p ) =  \sum_{p=1}^{2s} \r{log}\, ([k_{p-1},k_{p-2}, k_{p-3}, \dots])
\cdot [k_p,k_{p+1},k_{p+2},\dots ]).
\eqno(2.4)
$$
On the other hand, according to the formula (3.2.1) in [Se85], we have
$$
 \r{dist}\, (x_0,x_1)= \frac{1}{2}\r{log}([k_0,k_{-1},k_{-2},\dots ]\cdot [k_1,k_2,\dots ]\cdot
 [k_1,k_0,k_{-1},\dots ]\cdot [k_2,k_3,\dots ])
 $$
and hence, more generally,
$$
 \r{dist}\, (x_{2r},x_{2r+1})= 
 $$
 $$
\frac{1}{2} \r{log}([k_{2r},k_{2r-1},k_{2r-2},\dots ]\cdot [k_{2r+1},k_{2r+2},\dots ]\cdot
 [k_{2r+1},k_{2r},k_{2r-1},\dots ]\cdot [k_{2r+2},k_{2r+3},\dots ]).
 \eqno(2.5)
 $$
Inserting  (2.5) into the  r.h.s. of (2.3), we will see that it agrees with the r.h.s. of (2.4).
This completes the proof.

\medskip

The formula (2.3)  justifies identification of distance measured along a geodesic with
(doubly) logarithmic cosmological time in the next section.

\smallskip
During the stretch of {\it time/geodesic length} which such a geodesic spends
in the vicinity of a   vertex of $\Delta$, the respective space--time in a certain sense
can be approximated by its degenerate version, corresponding to the vertex itself,
and this justifies considering the respective
segments of geodesics  as the
``instanton Kasner eras''.

\bigskip

\centerline{3. \bf Mixmaster chaos in complex time and Painlev\'e VI}

\medskip

{\bf 3.1.  Painlev\'e VI.}    Contrary to the separatrix approximation methods,
 the results about encoding of geodesics $\beta$ with irrational ends and 
formulas for the distances between consecutive cutting points are exact, but
we did not yet introduce  analogs of space--times fibered over geodesics as 
theirs time axes.  We will do it in this section. The respective space--times 
are (complexified) versions of Bianchi IX models with $SU(2)$ (rather than $SO(3)$)
action, the so called gravitational instantons. 
An important class of them is described by solutions of the  Painlev\'e VI equation
corresponding to a particular point in the space of parameters of these equations:
for us, the main references will be [To94], [Hi95], and [BaKo98].

\smallskip

However, the hyperbolic billiard's picture of sec.~2 can be lifted
to essentially arbitrary Painlev\'e VI equations, and we will start this section
with a brief explanation of  the relevant  formalism.

\smallskip

Equations of the type  Painlev\'e VI
form a four--parametric family. If the parameters $({\alpha ,\beta , \gamma ,\delta})$ are chosen,
the corresponding equation for a function $X(t)$ looks as follows:
$$
\frac{d^2X}{dt^2}=\frac{1}{2}\left(
\frac{1}{X}+\frac{1}{X-1}+\frac{1}{X-t}\right)
\left(\frac{dX}{dt}\right)^2 -
\left(
\frac{1}{t}+\frac{1}{t-1}+\frac{1}{X-t}\right)\frac{dX}{dt} +
$$
$$
+\frac{X(X-1)(X-t)}{t^2(t-1)^2}
\left[ \alpha +
\beta\frac{t}{X^2}+\gamma\frac{t-1}{(X-1)^2}+
\delta\frac{t(t-1)}{(X-t)^2}\right]. \eqno{(3.1)}
$$
In 1907, R.~Fuchs has rewritten (3.1) in the form $$
t(1-t)\left[t(1-t)\frac{d^2}{dt^2}+(1-2t)\frac{d}{dt}-\frac{1}{4}\right]
\int_{\infty}^{(X,Y)}\frac{dx}{\sqrt{x(x-1)(x-t)}}=
$$
$$
=\alpha Y+\beta\frac{tY}{X^2}+\gamma\frac{(t-1)Y}{(X-1)^2}
+(\delta -\frac{1}{2})\frac{t(t-1)Y}{(X-t)^2}  \eqno{(3.2)}
$$
Here he enhanced $X:=X(t)$ to $(X,Y):=(X(t),Y(t))$ treating the latter pair as
 a  section $P:= (X(t),Y(t))$ of the generic elliptic curve
$E =E(t):\ Y^2=X(X-1)(X-t)$. The section can be  local and/or multivalued.
\smallskip

In this form, the left hand side of (3.2) which we denote $\mu ( P)$ has a beautiful property: 
it is a non--linear differential expression (additive differential character) in coordinates of $P$
such that $\mu (P+Q)=\mu (P) +\mu (Q)$ where $P+Q$ means addition
of points of the generic elliptic curve $E$, with infinite section as zero. 
 In particular,
$\mu (Q) =0$ for points of finite order. 

\smallskip
To see it, notice that the integral in the l.h.s.
of (3.2) is additive modulo periods of our elliptic  curve, considered
as multivalued functions of $t$. These periods
 are annihilated by the Gauss differential operator
which is put before the integral sign in (3.2).
\smallskip

The right hand side of (3.2) looks more mysterious. In order to clarify its meaning, notice that
 $\mu (P)$  is defined up to multiplication by an invertible function of $t$.
\smallskip
If we choose a differential of the first kind $\omega$ on the generic curve and 
 the symbol of the Picard--Fuchs operator of the second order
annihilating periods of $\omega$,  the character will be defined uniquely.
Moreover, it is functorial with respect to base changes (cf. [Ma96], sec.~0.2, 1.2, 1.3).
In particular, if we pass to the analytic picture replacing the algebraic family
of curves $E(t)$ by the analytic  one $E_{\tau}:= \bold{C}/(\bold{Z}+\bold{Z}\tau )\mapsto \tau\in H$,
and denote by $z$ a fixed coordinate on $\bold{C}$, then (3.1) and (3.2) can be equivalently written
in the form
$$
\frac{d^2z}{d\tau^2}=\frac{1}{(2\pi i)^2}\sum_{j=0}^3
\alpha_j\wp_z(z+\frac{T_j}{2},\tau )
\eqno{(3.3)}
$$
where $(\alpha_0,\dots ,\alpha_3):=(\alpha ,-\beta ,\gamma ,\frac{1}{2}-\delta )$ and
$(T_0, T_1, T_2,T_3):=(0,1,\tau ,1+\tau )$, and
$$
\wp (z,\tau ):= \frac{1}{z^2}+\sum_{(m,n)\ne (0,0)}
\left(\frac{1}{(z-m\tau -n)^2}-\frac{1}{(m\tau +n)^2}\right) .     
\eqno{(3.4)}
$$
Moreover, we have
$$
\wp_z (z,\tau )^2=4(\wp (z,\tau )-e_1(\tau ))(\wp (z,\tau)-e_2(\tau ))
(\wp (z,\tau )-e_3(\tau ))
\eqno{(3.5)}
$$
where
$$
e_i(\tau )=\wp (\frac{T_i}{2},\tau ),
 \eqno{(3.6)}
$$
so that $e_1+e_2+e_3=0$.

\smallskip

The family Painlev\'e VI was written in this form in [Ma06]. It was
considerably generalised by K.~Takasaki in [Ta01], in particular,
he found its versions for other families of Painlev\'e equations.

\smallskip

Now, any multivalued solution $z=z(\tau )$ of (3.3) defines a multi--section of
the family which is a covering of $H$. 
 In particular, if we can control its ramification  and monodromy, then we may consider
its behavior over geodesics with ideal ends in $H$ and study the relevant statistical properties.
The most accessible examples are  algebraic solutions classified in [Boa08], [LiTy08] and other
works.

\smallskip

However, here we will return to Bianchi IX models, which according to
[Hi98] correspond to the equation with parameters $(\alpha ,\beta ,\gamma , \delta)=
(\frac{1}{8},  -\frac{1}{8}, \frac{1}{8},  \frac{3}{8})$, solvable in elliptic
functions. We will skip the beautiful twistor geometry bridging Painlev\'e VI
and Bianchi IX and simply reproduce the relevant results from
[To94] and [Hi95], somewhat reworked and simplified in [BaKo98]. 

\medskip

{\bf 3.2. $SU(2)$  Bianchi IX metric and scaling factors.} 
Consider the $SU(2)$ Bianchi IX model with metric of the form
$$
g = F\left( d\mu^2 +\frac{\sigma_1^2}{W_1^2} +\frac{\sigma_2^2}{W_2^2}
+\frac{\sigma_3^2}{W_3^2}\right) .
\eqno(3.7)
$$
Here $\mu$ is cosmological time,  $(\sigma_j)$ are $SU(2)$--invariant 
forms along space--sections with $d\sigma_i=\sigma_j\wedge \sigma_k$
for all cyclic permutations of $(1,2,3)$, and $F$ is a conformal factor.

\smallskip

By analogy with the $SO(3)$ case and metric $dt^2-a(t)^2dx^2-b(t)^2dy^2-c(t)^2dz^2$, 
we may and  will treat $W_i$ (as well as some natural monomials in $W_i$ and $F$) as
{\it $SU(2)$--scaling factors.}

\smallskip

However, contrary to the $SO(3)$--case,  generic solutions of Einstein equations
in the $SU(2)$--case can be written explicitly in terms of elliptic modular functions, 
whereas their chaotic behaviour
along geodesics in the complex half--plane of time is  only a reflection
of the chaotic behaviour of the respective billiard ball trajectories.

\medskip

{\bf 3.3. Theta--functions with characteristics.} Explicit formulas in [BaKo98]
use the following basic function of the complex arguments $i\mu\in H$, $z\in \bold{C}$,  with
parameters  $(p,q)$ called theta--characteristics:
$$
\vartheta [p,q](z, i\mu ):=\sum_{m\in \bold{Z}} \roman{exp} \{ -\pi  (m+p)^2\mu
+ 2\pi i (m+p)(z+q)\}.
\eqno(3.8)
$$
It can be expressed through the theta--function with vanishing characteristics:
$$
\vartheta [p,q](z, i\mu ) = \roman{exp}\,\{-\pi p^2\mu+2\pi i pq\}\cdot 
\vartheta [0,0](z +pi\mu + q,  i\mu ) .
\eqno(3.9)
$$
All these functions satisfy classical automorphy identities with respect
to the action of $PGL(2,\bold{Z})$.

\medskip

{\bf 3.4. Theorem.} ([To94], [Hi95], [BaKo98].) {\it  Put
$$
\vartheta [p,q]:= \vartheta [p,q] (0,i\mu )
\eqno(3.10)
$$
and
$$
\vartheta_2:= \vartheta [1/2,0],\ 
\vartheta_3:= \vartheta [0,0],\
\vartheta_4:=  \vartheta [0,1/2].
\eqno(3.11)
$$
\smallskip
 (A) Consider the following
scaling factors as functions of $\mu$ with parameters $(p,q)$:
$$
W_1:=-\frac{i}{2}\vartheta_3\vartheta_4 \frac{\frac{\d}{\d q} \vartheta [p,q+1/2]}{e^{\pi ip} \vartheta [p,q]},\
W_2:=\frac{i}{2}\vartheta_2\vartheta_4 \frac{\frac{\d}{\d q} \vartheta [p+1/2,q+1/2]}{e^{\pi ip} \vartheta [p,q]},
$$
$$
W_3:=-\frac{1}{2}\vartheta_2\vartheta_3 \frac{\frac{\d}{\d q} \vartheta [p+1/2,q]}{ \vartheta [p,q]},
\eqno(3.12)
$$
Moreover, define the conformal factor $F$ with non--zero cosmological constant $\Lambda$ by
$$
F:= \frac{2}{\pi\Lambda} \frac{W_1W_2W_3}{( \frac{\d}{\d q} \roman{log}\, \vartheta [p,q] )^2}
\eqno(3.13)
$$
The metric (3.7) with these scaling factors for real $\mu >0$ is real and satisfies
the Einstein equations if either
$$
\Lambda < 0,\ p\in \bold{R},\   q\in \frac{1}{2}+ i \bold{R},
\eqno(3.14)
$$
or
$$
\Lambda > 0,\ q\in \bold{R},\   p\in \frac{1}{2}+ i \bold{R} .
\eqno(3.15)
$$

\medskip
(B)  Consider now a different system of scaling factors
$$
W_1^{\prime}:= \frac{1}{\mu+q_0} + 2 \frac {d}{d\mu} \roman{log}\,\vartheta_2,\
W_2^{\prime}:= \frac{1}{\mu+q_0} + 2 \frac {d}{d\mu} \roman{log}\,\vartheta_3,\
$$
$$
W_3^{\prime}:= \frac{1}{\mu+q_0} + 2 \frac {d}{d\mu} \roman{log}\,\vartheta_4,\
\eqno(3.16)
$$
and
$$
F^{\prime}:= C(\mu +q_0)^2\,W_1^{\prime}W_2^{\prime}W_3^{\prime},
\eqno(3.17)
$$
where $q_0, C\in \bold{R}$, $C>0$.
\smallskip
The metric (3.7) with these scaling factors for real $\mu>0$  is real and satisfies
the Einstein equations with vanishing cosmological constant.}

\medskip

 We will now consider
values of $i\mu \in \Delta \subset \overline{H}$ in the vicinity
of $i\infty$ but not necessarily lying on the imaginary axis.
Since we are interested in the instanton analogs of Kasner's solutions,
we will collect basic facts about asymptotics of scaling factors
for $i\mu \to i\infty$.

\smallskip
For brevity, we will call a number $r\in \bold{R}$ {\it general},  if
$r\notin \bold{Z}\cup (1/2+\bold{Z})$.

\smallskip

For such $r$, denote by $\langle r\rangle \in (-1/2,0) \cup (0,1/2)$ such real number
that $r+m_0=\langle r \rangle$ for a certain (unique) $m_0\in \bold{Z}$.

\medskip

{\bf 3.5. Theorem.} {\it The scaling factors of the Bianchi IX spaces listed in
Theorem 3.4 have the following asymptotics near $\mu =+\infty$:
\smallskip

(i) For $\Lambda =0$:
$$
W_1^{\prime} \sim \frac{\pi}{2},\quad
W_2^{\prime} \sim
W_3^{\prime} \sim \frac{1}{\mu +q_0} .
\eqno(3.18)
$$

\smallskip

(ii) For $\Lambda <0$ and general $p$:
$$
W_1\sim \pi \langle p\rangle\,\roman{exp}\, \{\pi i(\langle p\rangle -p)\},\quad 
W_2\sim \pm W_3,
$$
$$
W_3 \sim -2\pi i \, \langle p+1/2\rangle \cdot \roman{exp}\,\{ \pi i\, \roman{sgn}\,\langle p\rangle q\}
\cdot 
\roman{exp} \{\pi\mu (|\langle p\rangle|- 1/2)\}.
\eqno(3.19)
$$
\smallskip
(iii) For $\Lambda >0$, real $q$ and $p=1/2+ip_0, p_0\in \bold{R}$:
$$
-W_1 \sim \pi p_0\,\roman{tan} \{\pi (q-p_0\mu)\} - \frac{1}{2},
\quad  W_2 \sim -W_3, \quad 
$$
$$
W_3\sim 2\pi p_0\cdot (\roman{cos}\,\pi(q- p_0\mu ))^{-1}.
\eqno(3.20)
$$
}

\medskip

{\it Comments.} Theorem 3.5 shows that for general members of all
solution families from [BaKo98], after eventual sign changes of some  $W_i$'s
and outside of the pole singularities on the real time axis,
we have asymptotically  $W_2=W_3$, $W_1\ne W_2$. 
\smallskip

In the next section, we will show that  such  condition, when it is satisfied
exactly rather than asymptotically, allows
one to quantize the respective geometric picture in terms of Connes--Landi
([CoLa01]. This gives additional substance to our vision
that chaotic Mixmaster evolution  along hyperbolic geodesics 
reflects a certain ``dequantization'' of the hot quantum early Universe.
\smallskip

Sign changes alluded to above
are allowed, since Babich and Korotkin get their much simpler formulas
by cleverly extracting square roots from expressions given in [Hi95]. Moreover,
in the second version of their paper posted in arXiv in March 2014, they corrected
the signs of $W_1$ and $C$ (cf. their Lemma 2). For our purposes, this is not
essential.

\medskip

{\bf Proof of Theorem 3.5.}
Directly from (3.9)--(3.11), we obtain:
$$
\vartheta_2=\sum_{m\in \bold{Z}} \roman{exp} \{-\pi(m+\frac{1}{2})^2\mu\}\  \sim
2\,\roman{exp}\{-\pi\mu/4\}  , 
\eqno(3.21)
$$
$$
\vartheta_3=\sum_{m\in \bold{Z}} \roman{exp} \{-\pi m^2\mu\}\ \sim 1+2\,\roman{exp}\{-\pi\mu\}  ,
\eqno(3.22)
$$
$$
\vartheta_4=\sum_{m\in \bold{Z}} \roman{exp} \{-\pi m^2\mu \} (-1)^m  \ \sim 1-2\,\roman{exp}\{-\pi\mu\}  .
\eqno(3.23)
$$
\smallskip
Therefore
$$
 \frac {d}{d\mu} \roman{log}\,\vartheta_2 \sim -\frac{\pi}{4},\quad 
 \frac {d}{d\mu} \roman{log}\,\vartheta_3 \sim -2\pi\,  \roman{exp}\{-\pi\mu\},\quad
\frac {d}{d\mu} \roman{log}\,\vartheta_4 \sim  2\pi \, \roman{exp}\{-\pi\mu\} .
$$
From this and (3.16), (3.17) one gets (3.18) for $\Lambda = 0.$
\smallskip

Now consider the case $\Lambda < 0$, $p$ general.

\smallskip

Then from (3.8), (3.10),  and (3.14) one gets
$$
\vartheta [p,q] =  \sum_{m\in \bold{Z}} \roman{exp} \{ -\pi  (m+p)^2\mu
+ 2\pi i (m+p)q\} \sim
$$
$$
\sim \roman{exp} \{2\pi i \langle p \rangle q\}\cdot \roman{exp}\{-\pi \langle p \rangle^2\mu\},
\eqno(3.24)
$$
because for general $p$, the leading term of $\vartheta [p,q]$ corresponds to the unique
value of $m$ for which $(m+p)^2$ is minimal, that is, equals $\langle p \rangle^2$.
\smallskip

Hence
$$
\frac{\d}{\d q}\vartheta [p,q] \sim  2\pi i \, \langle p\rangle 
\roman{exp} \{2\pi i \langle p \rangle q\}\cdot \roman{exp}\{-\pi \langle p \rangle^2\mu\} .
\eqno(3.25)
$$
Thus, from (3.12), and (3.21)--(3.25) we obtain
$$
-W_1=\frac{i}{2}\vartheta_3\vartheta_4 \frac{\frac{\d}{\d q} \vartheta [p,q+1/2]}{e^{\pi ip} \vartheta [p,q]}\sim
\frac{i}{2}\cdot 2\pi i \, \langle p\rangle 
\roman{exp} \{2\pi i \langle p \rangle (q+1/2)\}\cdot \roman{exp}\{-\pi \langle p \rangle^2\mu\} \times
$$
$$
 \roman{exp}\,\{-\pi i p\}\cdot \roman{exp} \{-2\pi i \langle p \rangle q\}\cdot \roman{exp}\{\pi \langle p \rangle^2\mu\} 
 = -\pi \langle p\rangle\,\roman{exp}\, \{\pi i(\langle p \rangle -p)\} .
 $$
 
 Furthermore,
$$
W_2=\frac{i}{2}\vartheta_2\vartheta_4 \frac{\frac{\d}{\d q} \vartheta [p+1/2,q+1/2]}{e^{\pi ip} \vartheta [p,q]}\sim
$$
$$
\sim \frac{i}{2}\cdot 2\,\roman{exp}\{-\pi\mu/4\}\cdot
2\pi i \, \langle p+1/2\rangle \cdot
\roman{exp} \{2\pi i \langle p+1/2 \rangle (q+1/2)\}\cdot \roman{exp}\{-\pi \langle p+1/2 \rangle^2\mu\} \times
$$
$$
 \roman{exp}\,\{-\pi i p\}\cdot \roman{exp} \{-2\pi i \langle p \rangle q\}\cdot \roman{exp}\{\pi \langle p \rangle^2\mu\}\sim
 $$
$$
-2\pi\, \langle p+1/2\rangle \roman{exp}\,
\{\pi i [\langle p+1/2 \rangle - p -\roman{sgn} \langle p \rangle q]\} \cdot 
\roman{exp} \{\pi\mu (|\langle p\rangle|- 1/2)\}.
$$

\bigskip

Notice that  exponential terms were rewritten using the identity
$$
\langle p+1/2\rangle = \langle p\rangle - \frac{1}{2} \roman{sgn} \langle p\rangle .
$$

\smallskip
Similarly,
$$
W_3:=-\frac{1}{2}\vartheta_2\vartheta_3 \frac{\frac{\d}{\d q} \vartheta [p+1/2,q]}{ \vartheta [p,q]}\sim
$$
$$
\sim -\frac{1}{2}\cdot 2\,\roman{exp}\{-\pi\mu/4\}\cdot
2\pi i \, \langle p+1/2\rangle \cdot
\roman{exp} \{2\pi i \langle p+1/2 \rangle q\}\cdot \roman{exp}\{-\pi \langle p+1/2 \rangle^2\mu\} \times
$$
$$
\roman{exp} \{-2\pi i \langle p \rangle q\}\cdot \roman{exp}\{\pi \langle p \rangle^2\mu\}\sim
$$
$$
-2\pi i \, \langle p+1/2\rangle \cdot \roman{exp}\,\{ \pi i\, \roman{sgn}\,\langle p\rangle q\}
\cdot \roman{exp} \{\pi\mu (|\langle p\rangle|- 1/2)\}.
$$
\smallskip

Comparing expressions for $W_2$ and $W_3$, one easily sees that $W_2=\pm W_3$,
where the exact sign can be expressed through $p$ and $q$.
\smallskip
For the conformal factor (3.13) we then get the following asymptotic:
$$
F= \frac{2}{\pi\Lambda} \frac{W_1W_2W_3}{( \frac{\d}{\d q} \roman{log}\, \vartheta [p,q] )^2} \sim
$$
$$
2i \frac{\langle p+1/2\rangle^2}{\Lambda \langle p \rangle^2} \cdot
\roman{exp}\, \{ \langle p+1/2 \rangle + \langle p \rangle +2\,\roman{sgn}\,\langle p\rangle\,q\} \cdot 
\roman{exp} \{\pi \mu\, (2\,| \langle p \rangle |-1) \}.
$$

\medskip

Finally, pass to the case $\Lambda >0$. Put $p= \frac{1}{2} + ip_0,\ p_0\in \bold{R}$.
We have again to locate first the leading terms as $\mu \to +\infty$ in
$$
\vartheta [p,q] =  \sum_{m\in \bold{Z}} \roman{exp} \{ -\pi  (m+p)^2\mu
+ 2\pi i (m+p)q\} ,
$$
and also respective terms when $p$ and/or $q$ are shifted by $1/2$.
Obviously, they correspond to the minimal values of
$\roman{Re}\, (m+p)^2$, resp.  $\roman{Re}\, (m+p+1/2)^2$, for $m\in \bold{Z}$.
Since
$$
\roman{Re}\, (m+p)^2 = (m+\frac{1}{2})^2-p_0^2,\quad
\roman{Re}\, (m+p+\frac{1}{2})^2 = (m+1)^2-p_0^2,
$$
in the first case there are two leading terms, for $m=0$ and $m=-1$,
and in the second case just one, for $m=-1$.
\smallskip

Thus, for $\Lambda > 0$, we have
$$
\vartheta [p,q] \sim   \roman{exp} \{ \pi \mu (p_0^2-1/4)\}\cdot [  \roman{exp} \{2\pi ipq -\pi i p_0\mu\} 
+\roman{exp} \{2\pi i (p-1)q + \pi i p_0\mu \} ].
$$

The sum of two terms in square brackets can be rewritten
so that in the end we obtain
$$
\vartheta [p,q] \sim   \roman{exp} \{ \pi \mu (p_0^2-1/4)\}\cdot \roman{exp}\,\{-2\pi p_0q\}
\cdot \roman{cos}\,\pi(q- p_0\mu ).
\eqno(3.26)
$$
$$
\vartheta [p+1/2,q] \sim   \roman{exp} \{ \pi \mu p_0^2\} \cdot  \roman{exp} \{-2\pi p_0q\}.
\eqno(3.27)
$$
When we have to replace a real $q$ by $q+1/2$, we may do it formally in the right hand side expressions 
in (3.26), (3.27).

\smallskip

Therefore, we have from (3.12), (3.22) and (3.27):
$$
\frac{W_2}{W_3} =-i\cdot \frac{\vartheta_4}{\vartheta_3} \cdot
\frac{\frac{\partial}{\partial q}\vartheta [p+1/2,q+1/2]}{e^{\pi i p}\frac{\partial}{\partial q} \vartheta [p+1/2,q]}\sim
$$
$$
i\cdot \frac{\roman{exp} \{-2\pi p_0(q+1/2)\} } {\roman{exp} \{\pi i (1/2 +ip_0)\}\cdot\roman{exp} \{-2\pi p_0 q\} } =
-1.
$$
Now,
$$
W_3:=-\frac{1}{2}\vartheta_2\vartheta_3 \frac{\frac{\d}{\d q} \vartheta [p+1/2,q]}{ \vartheta [p,q]}\sim
$$
$$
 \frac{\roman{exp}\{-\pi\mu/4\}\cdot (2\pi p_0)\cdot     \roman{exp} \{ \pi \mu p_0^2\} \cdot  \roman{exp} \{-2\pi p_0q\}}{\roman{exp} \{ \pi \mu (p_0^2-1/4)\} \cdot \roman{exp}\,\{-2\pi p_0q\} \cdot \roman{cos}\,\pi(q- p_0\mu )} \sim
$$
$$
2\pi p_0\cdot (\roman{cos}\,\pi(q- p_0\mu ))^{-1}.
$$
Furthermore,
$$
-W_1=\frac{i}{2}\vartheta_3\vartheta_4 \frac{\frac{\d}{\d q} \vartheta [p,q+1/2]}{e^{\pi ip} \vartheta [p,q]}\sim
$$
$$
\frac{i}{2}\cdot \frac{ \roman{exp} \{ \pi \mu (p_0^2-1/4)\}\cdot \frac{\d}{\d q} [\roman{exp}\,\{-2\pi p_0(q+1/2)   \}
\cdot \roman{cos}\,\pi(q+ 1/2 - p_0\mu )]} {\roman{exp}\, \{\pi (i/2 - p_0)\} \cdot
 \roman{exp} \{ \pi \mu (p_0^2-1/4)\}\cdot \roman{exp}\,\{-2\pi p_0q\}\cdot \roman{cos}\,\pi(q- p_0\mu )}\sim
 $$
 $$
\frac{i}{2}\cdot \frac{\frac{\d}{\d q} [\roman{exp}\,\{-2\pi p_0(q+1/2)   \}
\cdot \roman{cos}\,\pi(q+ 1/2 - p_0\mu )]} {\roman{exp}\, \{\pi (i/2 - p_0)\} 
 \cdot \roman{exp}\,\{-2\pi p_0q\}\cdot \roman{cos}\,\pi(q- p_0\mu )} \sim
 $$
 $$
- \frac{1}{2} \cdot \frac{\frac{\d}{\d q} [\roman{exp}\,\{-2\pi p_0(q+1/2)   \}
\cdot \roman{sin}\,\pi(q - p_0\mu )]} { 
 \roman{exp}\,\{-2\pi p_0(q+ 1/2)\}\cdot \roman{cos}\,\pi(q- p_0\mu )}  \sim
 $$
$$
\pi p_0\,\roman{tan} \{\pi (q-p_0\mu)\} - \frac{1}{2}.
$$
This completes the proof of Theorem 3.5.

\bigskip

\centerline{4. \bf Theta deformations of gravitational instantons}

\medskip

{\bf 4.1. Theta deformations.} In Section 5 of [MaMar14] we showed that the gluing of space--times
across the singularity using an algebro-geometric blowup
can be made compatible with the idea of spacetime coordinates becoming
noncommutative in a neighborhood of the initial singularity where quantum
gravity effects begin to dominate. 

\smallskip

This compatibility is described there
 in terms of Connes--Landi theta deformations ([CoLa01]) and
Cirio--Landi--Szabo toric deformations ([CiLaSz13])  of Grassmannians.

\smallskip

Here we consider the same problem in the case of the Bianchi IX models
with $SU(2)$-symmetry, namely whether they can be made
compatible with the hypothesis of noncommutativity at the Planck scale, using
isospectral theta deformations. 

\smallskip
The metrics on the $S^3$ sections, in this case,
are only left $SU(2)$--invariant. We show that among all the $SU(2)$ Bianchi
IX spacetime, the only ones that admit isospectral theta--deformations of their 
spatial $S^3$--sections are those where the metric tensor
$$
g = w_1 w_2 w_3\, d\mu^2 + \frac{w_2 w_3}{w_1}\, \sigma_1^2 + \frac{w_1w_3}{w_2}\, \sigma_2^2
+\frac{w_1 w_2}{w_3}\, \sigma_3^2 
\eqno(4.1)
$$
is of the special form satisfying $w_1\neq w_2 = w_3$ (the two directions $\sigma_2$ and
$\sigma_3$ have equal magnitude). In these metrics, the $S^3$ sections are Berger spheres. 
This class includes the general Taub-NUT family ([Taub51], [NUT63]), and the 
Eguchi--Hanson metrics ([EgHa79a], [EgHa79b]). The theta--deformations are obtained, 
as in the case of the deformations $S^3_\theta$ of [CoLa01] of the round $3$-sphere,
by deforming all the tori of the Hopf fibration to noncommutative tori.  

\medskip

{\bf 4.2. Proposition.} {\it A Bianchi IX Euclidean spacetime $X$ with $SU(2)$--symmetry 
admits a noncommutative theta-deformation $X_\theta$, obtained by deforming
the tori of the Hopf fibration of each spacial section $S^3$ to noncommutative tori,
if and only if its metric has the $SU(2)\times U(1)$--symmetric form
$$
g = w_1w_3^2\, d\mu^2 +\frac{w_3^2}{w_1} \, \sigma_1^2 + w_1 \, (\sigma_2^2 +\sigma_3^2)  .
\eqno(4.2)
$$
}

\medskip

{\bf Proof.} In appropriate local coordinates  the $SU(2)$--invariant forms $(\sigma_i)$ satisfying relations
$d\sigma_i = \sigma_j \wedge \sigma_k$ for all cyclic permutations $(i,j,k)$ have the explicit form 
$$
 \sigma_1 = x_1 \, dx_2 - x_2\, dx_1 + x_3\, dx_0 - x_0 \, dx_3   = 
\frac{1}{2} (d\psi +\cos\theta \, d\phi), 
$$
$$ 
\sigma_2 = x_2 \, dx_3 - x_3 dx_2 + x_1 \, dx_0 - x_0\, dx_1   = 
\frac{1}{2} (\sin \psi \, d\theta - \sin\theta \cos \psi \, d\phi), 
$$
$$ 
\sigma_3 = x_3 \, dx_1 - x_1 \, dx_3+ x_2 \, dx_0 - x_0 \, dx_2  = 
\frac{1}{2} (-\cos \psi\, d\theta - \sin \theta\, \sin\psi\, d\phi), 
$$
with Euler angles $0\leq \theta \leq \pi$, $0\leq \phi \leq 2\pi$ and $0\leq \psi \leq 4\pi$ (for the $SU(2)$
case). 
\smallskip
The Hopf coordinates $(\xi_1, \xi_2,\eta )$ are defined by
$$ 
z_1:= x_1+ix_2 = e^{i (\psi+\phi)} \cos \frac{\theta}{2}  = e^{i\xi_1} \cos\eta, 
$$
$$ 
z_2 :=x_3+ix_0 = e^{i (\psi-\phi)} \sin \frac{\theta}{2}  = e^{i\xi_2} \sin\eta . 
$$

 Equivalently, identifying $S^3$ with unit
quaternions, we write $q\in SU(2)$ as
$$
 q :=  \pmatrix z_1 & z_2 \\ -\bar z_2 & \bar z_1 \endpmatrix  = \pmatrix e^{i\xi_1}\cos\eta & 
e^{i\xi_2} \sin\eta \\ -e^{-i\xi_2}\sin\eta & e^{-i\xi_1}\cos\eta \endpmatrix, 
$$
where $|z_1|^2+|z_2|^2=1$ and $(\xi_1,\xi_2,\eta)$ are the Hopf coordinates as above. 
\smallskip
The noncommutative $\theta$--deformations ([CoLa01]) of the $3$--sphere $S^3$ are
obtained by deforming all the $2$--tori of the Hopf fibration to noncommutative
$2$--tori $T^2_\theta$. Namely, replace $q$ with
$$
 \pmatrix U \, \cos\eta & 
V\, \sin\eta \\ - V^* \, \sin\eta & U^*\, \cos\eta \endpmatrix, 
$$
where $U,V$ are the generators of the noncommutative $2$--torus $T^2_\theta$.

\smallskip
Then one obtains the algebra generated by $\alpha= U \cos\eta$ and
$\beta = V \sin\eta$ with $\alpha\beta = e^{2\pi i \theta} \beta\alpha$,
$\alpha^*\beta =e^{-2\pi i \theta} \beta\alpha^*$, $\alpha^* \alpha = \alpha \alpha^*$,
$\beta^*\beta=\beta\beta^*$ and $\alpha \alpha^*+ \beta \beta^* =1$.
It is shown in [CoLa01] that this deformation is {\it isospectral} with respect to the
bi-invariant round metric on $S^3$, in the sense that the data of the
Hilbert space of square integrable spinors $H=L^2(S^3 ,S)$ and the Dirac operator
$D$ for the round metric on $S^3$ give rise to spectral triples on the deformed
algebras $S^3_\theta$.
\smallskip
In fact, the general result of [CoLa01] shows that isospectral theta--deformations
can be constructed whenever there is an isometric torus action. In particular,
in our case the question reduces to whether the action of $T^2$ that rotates
the tori of the Hopf fibration preserves the Bianchi IX metric.
\smallskip
In Hopf coordinates the action of $T^2$ is given by $(t_1,t_2): (\xi_1,\xi_2) \mapsto (\xi_1 + t_1, \xi_2+t_2)$,
or in terms of the Euler angles, $(u,v): (\phi,\psi) \mapsto (\phi+u, \psi+v)$, with 
$t_1=(u+v)/2$ and $t_2=(v-u)/2$. It is immediate to check that the $U(1)$--action 
$u: \phi \mapsto \phi+u$ leaves the $1$-forms $\sigma_i$ invariant. This is the
$U(1)$-action of the Hopf fibration $S^1 \hookrightarrow S^3 \to S^2$. The form
$\sigma_1$ is also invariant under the other $U(1)$-action $v: \psi \mapsto \psi+v$,
while the other forms $\sigma_2, \sigma_3$ transform as
$$ v^* \sigma_2 = \frac{1}{2}(\sin (\psi+\beta)\, d\theta - \cos(\psi+\beta)\, \sin\theta\, d\phi) $$
$$ v^* \sigma_3 = \frac{1}{2} (-\cos(\psi+\beta)\, d\theta - \sin(\psi+\beta) \, \sin\theta \, d\phi),  $$
hence it is clear that we have $v^* g =g$ for a Bianchi IX metric
$$
g =  d\mu^2 + a^2\, \sigma_1^2 +b^2\, \sigma_2^2
+c^2\, \sigma_3^2 
\eqno(4.3)
$$
if and only if $b=c$. In the case $b=c$, with
$$ g =  d\mu^2 + \frac{a^2}{4} (d\psi +\cos\theta \, d\phi)^2 + \frac{c^2}{4} (d\theta^2 + \sin^2\theta\, d\phi^2), $$
the $T^2$ action is isometric and the
resulting theta-deformations are therefore isospectral, with spectral triples
$(A, H, D)$, with $A=C^\infty(S^3_\theta)$, and spinors $H=L^2(S^3,S)$ and 
Dirac operator $D$ with respect to the Bianchi IX metric with $b=c$.

\smallskip

This is in stark contrast with the situation described in [EsMar13], where (Lorentzian
and Euclidean) Mixmaster cosmologies of the form
$$ 
\mp dt^2 + a(t)^2 dx^2 + b(t)^2 dy^2 + c(t)^2 dz^2 
$$
were considered, with $T^3$-spatial sections, which always admit isospectral
theta-deformations.

\smallskip

We have recalled in the previous section how the self--duality equations
for the $SU(2)$ Bianchi IX models can be described in terms of Painlev\'e
VI equations [To94], [Hi95], [Ok98], and how the general solutions 
(with $w_1\neq w_2 \neq w_3$) can be written explicitly in terms of 
theta constants [BaKo98], and are obtained from a Darboux--Halphen 
type system [PeVa12], [Tak92]. In the case of the family of Bianchi IX models 
with $SU(2)\times U(1)$-symmetry, considered in Proposition 42, this
system has algebraic solutions that give 
$$ 
w_2 = w_3 = \frac{1}{\mu - \mu_0}, \ \ \  w_1 = \frac{\mu-\mu_*}{(\mu-\mu_0)^2}, 
\eqno(4.4)
$$
with singularities at $\mu_*$ (curvature singularity), $\mu_0$ (Taubian infinity) and $\infty$ (nut). 
The condition $\mu_* < \mu_0$ avoids naked singularities, by hiding the curvature singularity 
at $\mu_*$ behind the Taubian infinity, see the discussion in Section 5.2 of [PeVa12]. 

\medskip

Consider the operator
$$ 
D_B = -i  \pmatrix \frac{1}{\lambda}X_1 & X_2 +i X_3 \\ X_2-iX_3 
& -\frac{1}{\lambda}X_1 \endpmatrix + \frac{\lambda^2 +2}{2\lambda}, 
\eqno(4.5) 
$$
where $\{ X_1, X_2, X_3 \}$ constitute a basis of the Lie algebra orthogonal for
the bi--invariant metric. Assume moreover that the left--invariant metric on $S^3$ is diagonal
in this basis, with
eigenvalues $\{ w^2/w_1, w_1, w_1 \}$, with $w=w_2=w_3$ and $\lambda=w/w_1$,
and where the $w_i$ are as in (4.4). Consider also the operator 
$$ D =  \frac{1}{w_1^{1/2} w} \left( \gamma^0 
\left( \frac{\partial}{\partial \mu} + \frac{1}{2} ( \frac{\dot{w}}{w} + \frac{1}{2} \frac{\dot{w_1}}{w_1} ) \right)
+ w_1 \,\, D_B|_{\lambda = \frac{w}{w_1}} \right). 
\eqno(4.6) $$

\medskip

{\bf 4.2. Proposition. } {\it The operators $D$ of (4.6) give Dirac operators for
isospectral theta deformations of the $SU(2)\times U(1)$-symmetric spacetimes
of Proposition 4.2.}

\medskip

{\bf Proof.} We consider the frame $\theta^i$ with $i\in \{ 0,1,2,3 \}$, given by
$$ 
\theta^0= uw \, d\mu, \ \  \  \theta^1 = u \lambda\, \sigma_1, \ \ \ 
\theta^2 = u \, \sigma_2, \ \ \   \theta^3 =u \, \sigma_3, 
$$
where $u=w_1^{1/2}$ and $\lambda =w/w_1$, for $w=w_2=w_3$. Since the
$\sigma_i$ satisfy $d \sigma_i=\sigma_j \wedge \sigma_k$ for cyclic permulations
$\{ i,j,k \}$ of $\{1,2,3\}$, we have $d\theta^0=0$, and furthermore
$$ 
d\theta^1 = (\dot{u}\lambda+ u \dot{\lambda})\, d\mu \wedge \sigma_1 + 
u\lambda\,  \sigma_2 \wedge \sigma_3  = \frac{1}{uw} (\frac{\dot{u}}{u}+\frac{\dot{\lambda}}{\lambda}) \, \theta^0\wedge \theta^1 + \frac{1}{u\lambda} \,\, \lambda^2 \, \theta^2 \wedge \theta^3 ,
$$
$$
 d\theta^2 = \dot{u}\, d\mu \wedge \sigma_2 + u \sigma_3 \wedge \sigma_1 =
\frac{1}{uw}\, \frac{\dot{u}}{u} \theta^0\wedge \theta^2 + \frac{1}{u\lambda}\, \theta^3 \wedge \theta^1 ,
$$
$$ 
d\theta^3 = \dot{u}\, d\mu \wedge \sigma_3 + u \sigma_1 \wedge \sigma_2 =
\frac{1}{uw}\, \frac{\dot{u}}{u} \theta^0\wedge \theta^3 + \frac{1}{u\lambda}\, \theta^1 \wedge \theta^2
$$
where dot denotes the time derivative.
\smallskip
Proceeding then as in [ChCo12], we use the $d\theta^i$ to write the spin connection and we
obtain a Dirac operator of the form
$$ 
D = \gamma^0 \frac{1}{w_1^{1/2} w} 
\left( \frac{\partial}{\partial \mu} + \frac{1}{2}  (\frac{\dot{w}}{w} + \frac{1}{2} \frac{\dot{w_1}}{w_1}) \right)
+ \frac{w_1^{1/2}}{w} \,\, D_B|_{\lambda = \frac{w}{w_1}}, 
$$
or equivalently of the form (4.6), where $D_B$ is the Dirac operator on a Berger
$3$--sphere. The explicit form of Dirac operator on a Berger $3$--sphere  
with metric $\lambda^2 \sigma_1^2 +\sigma_2^2 +\sigma_3^2$ was computed in [Hi74],
and it is given by the operator (4.5). 

\smallskip

As in [EsMar13], the Dirac operator of Proposition 4.3 can be seen as involving 
an anisotropic Hubble parameter $H$. In the case of the metrics (4.3) of [EsMar]13 this was of the form
$$ H = \frac{1}{3} \left( \frac{\dot{a}}{a} + \frac{\dot{b}}{b} + \frac{\dot{c}}{c} \right) $$
with $a,b,c$ the scaling factors in (4.3). 

\smallskip

In the case of the $SU(2)$ Bianchi IX models, the anisotropic Hubble parameter is
again of the form $H = \frac{1}{3} ( H_1 + H_2 + H_3)$, where now 
the $H_i$ correspond to the three directions of the vectors dual to the
$SU(2)$-forms $\sigma_i$ in (4.1). For a metric of the form (4.2), or equivalently 
$$ g=  uw\, d\mu^2 + u^2 \lambda^2 \, \sigma_1^2 + u^2 \sigma_2^2 + u^2 \sigma_3^2, $$
with $u,\lambda,w$ as in Proposition 4.2, we take the anisotropic Hubble parameter to be 
$$ H =  \frac{1}{3} \left( \frac{\dot{u} \lambda + u \dot{\lambda}}{u\lambda} + 2 \frac{\dot{u}}{u}\right)
=  \frac{1}{3} \left( 3 \frac{\dot{u}}{u} + \frac{\dot{\lambda}}{\lambda} \right), $$
where
$$ \frac{\dot{u}}{u}  = \frac{1}{2} \frac{\dot{w_1}}{w_1}, \ \ \  \frac{\dot{\lambda}}{\lambda} =
\frac{\dot{w}}{w} - \frac{\dot{w_1}}{w_1}, $$
so that
$$ H =  \frac{1}{3} \left( \frac{\dot{w}}{w} + \frac{1}{2}  \frac{\dot{w_1}}{w_1} \right), $$
as in (4.6), so that we can write the 4-dimensional Dirac operator in the form 
$$ D= \gamma^0 \frac{1}{uw} \left( \frac{\partial}{\partial \mu} + \frac{3}{2} H \right) + D_B, $$
where $D_B =(w_1^{1/2}/w) \,\, D_B|_{\lambda = \frac{w}{w_1}}$ is the 
Dirac operator on the spatial sections $S^3$ with the left $SU(2)$-invariant
metric. 

\medskip

Notice that in the construction above we have considered the same modulus
$\theta$ for the noncommutative deformation of all the spatial sections $S^3$
of the Bianchi IX spacetime, but one could also consider a more general
situation where the parameter $\theta$ of the deformation is itself a function
of the cosmological time $\mu$. 
\smallskip

This would allow the dependence of
the noncommutativity parameter $\theta$ on the energy scale
(or on the cosmological timeline), with $\theta=0$ away from the singularity
where classical gravity dominates and noncommutativity only appearing near 
the singularity. Since a non--constant, continuously varying parameter $\theta$ crosses
rational and irrational values, this would give rise to a Hofstadter 
butterfly type picture, with both commutativity (up to Morita equivalence, as in
the rational noncommutative tori) and true noncommutativity (irrational
noncommutative tori). 

\medskip

Another interesting aspect of these noncommutative deformations
is the fact that, when we consider a geodesic in the
upper half plane encoding  Kasner eras in a mixmaster dynamics,
the points along the geodesic also determine a family of complex
structures on the noncommutative tori $T^2_\theta$ of the
theta--deformation of the respective spatial section.

\bigskip

\centerline{\bf References}
\medskip

[BaKo98] M.~V.~Babich, D.~A.~Korotkin. {\it Self--dual $SU(2)$--Invariant Einstein Metrics and Modular
Dependence of Theta--Functions.} Lett. Math. Phys. 46 (1998), 323--337. arXiv:gr-qc/9810025v2

\smallskip

[Boa08] Ph.~Boalch. {\it Towards a nonlinear Schwarz's list.} arXiv:0707.3375, 27 pp.
\smallskip

[Bo85]  O.~I.~Bogoyavlensky. {\it Methods in the qualitative theory of dynamical systems in astrophysics and gas dynamics.} 
Springer Series in Soviet Mathematics. Springer Verlag, Berlin, 1985. ix+301 pp.
\smallskip

[BoNo73] O.~I.~Bogoyavlenskii,  S.~P.~Novikov. {\it Singularities of the cosmological model
of the Bianchi IX type according to the qualitative theory of differential equations.}
Zh.~Eksp.~Teor.~Fiz.~64 (1973), 1475--1494.

\smallskip

[ChCo12] A.H.~Chamseddine, A.~Connes, {\it Spectral action for Robertson-Walker metrics}, 
J. High Energy Phys. (2012) N.10, 101, 29 pp.  

\smallskip

[CiLaSz13] L.S.~Cirio, G.~Landi, R.J.~Szabo, {\it Algebraic deformations of toric varieties. I. 
General constructions}, Adv. Math. 246 (2013), 33--88.

\smallskip
[CoLa01] A.~Connes, G.~Landi, {\it Noncommutative manifolds, the instanton algebra and isospectral deformations},
Comm. Math. Phys. 221 (2001), 141--159. 
\smallskip

[EgHa79a] T.~Eguchi, A.J.~Hanson, {\it Self-dual solutions to Euclidean Gravity}, Annals of Physics,
120 (1979), 82--106.

\smallskip

[EgHa79b] T.~Eguchi, A.J.~Hanson, {\it Gravitational Instantons}, Gen. Relativity Gravitation 
11, No 5 (1979) , 315--320.

\smallskip

[EsMar13] C.~Estrada, M.~Marcolli, {\it Noncommutative Mixmaster Cosmologies},
International Journal of Geometric Methods in Modern Physics 10 (2013) 1250086, 28 pp.

\smallskip
[Hi74] N.~Hitchin, {\it Harmonic spinors}, 
Advances in Math. 14 (1974), 1--55. 

\smallskip

[Hi95] N.~J.~Hitchin. {\it Twistor spaces, Einstein metrics and isomonodromic deformations.}
J.~Diff.~Geo., Vol.~42, No.~1 (1995), 30--112.

\smallskip

[KLKhShSi85]  I.~M.~Khalatnikov, E.~M.~Lifshitz, K.~M.~Khanin, L.~N.~Shchur, and Ya.~G.~Sinai.
{\it On the stochasticity in relativistic cosmology.} Journ. Stat. Phys., Vol.~38, Nos. 1/2 (1985), 97--114.

\smallskip

[LiTy08] O.~Lisovyy, Yu.~Tykhyy. {\it Algebraic solutions of the sixth Painlev\'e equation.}
arXiv:0809.4873

\smallskip 
[Ma96] Yu.~Manin. {\it Sixth Painlev\'e equation, universal elliptic curve,
and mirror of $\bold{P}^2$.}  In: geometry of Differential
Equations, ed. by A.~Khovanskii, A.~Varchenko, V.~Vassiliev.
Amer. Math. Soc.
Transl. (2), vol. 186  (1998), 131--151. arXiv:alg--geom/9605010.
\smallskip
[MaMar02]  Yu.~Manin, M.~Marcolli.  {\it Continued fractions, modular symbols, and non--commutative geometry.}
 Selecta math., new ser. 8 (2002),
475--521. 

arXiv:math.NT/0102006

\smallskip

[MaMar14] Yu.I.~Manin,  M.~Marcolli. {\it Big Bang, Blow Up, and Modular Curves: Algebraic Geometry in Cosmology.}
SIGMA Symmetry Integrability Geom. Methods Appl., 10 (2014), Paper 073, 20 pp. Preprint arXiv:1402.2158

\smallskip

[May87]   D.~Mayer.  {\it Relaxation properties of the Mixmaster Universe.} Phys. Lett. A, Vol. 121,
Nos. 8--9   (1987), 390--394.
[\smallskip

[NUT63] E.~Newman, L.~Tamburino, T.~Unti, {\it Empty-space generalization of the Schwarzschild metric}, 
Journ. Math. Phys. 4 (1963), 915--923.

\smallskip
[Ok98] S.~Okumura. {\it The self--dual Einstein--Weyl metric and classical solutions
of Painlev\'e VI.} Lett. in Math. Phys., 46 (1998), 219--232.

\smallskip

[PeVa12] P.M.~Petropoulos, P.~Vanhove, {\it Gravity, strings, modular and quasimodular forms},
Ann. Math. Blaise Pascal 19, No.~2 (2012),  379--430.

\smallskip

[Se85] C.~Series. {\it The modular surface and continued fractions.} J.~London MS, Vol. 2, no. 31  (1985), 69--80.

\smallskip

[Ta01] K.~Takasaki. {\it Painlev\'e--Calogero correspondence revisited.} Journ.~Math.~Phys., vol.~42, No 3 
(2001), 1443--1473.

\smallskip

[Tak92] L.~Takhtajan, {\it A simple example of modular forms as tau--functions for integrable equations}, 
Teoret. Mat. Fiz. 93 (1992), no. 2, 330--341. 

\smallskip

[Taub51] A.H.~Taub, {\it Empty space-times admitting a three parameter group of motions}, 
Annals of Mathematics 53 (1951), 472--490.

\smallskip

[To94] K.~P.~Tod. {\it Self--dual Einstein metrics from the Painlev\'e VI equation.} Phys.~Lett. A 190 (1994),
221--224.

\bigskip

\enddocument